\documentstyle[12pt]{article}
\begin{document}


\setlength{\oddsidemargin}{0 cm}
\setlength{\evensidemargin}{0 cm}
\setlength{\topmargin}{0 cm}
\setlength{\textheight}{22 cm}
\setlength{\textwidth}{16 cm}

\begin{flushright}
        INS-Rep.1043 \\
        July 1994 \\
        (revised)\\
\end{flushright}

\begin{center}
\begin{Large}
{\bf On the string equation at $c=1$}
\end{Large}

\noindent
  Toshio NAKATSU
\vspace{18pt}

\begin{small}
   {\it Institute for Nuclear Study,University of Tokyo} \\
   {\it Midori-cho,Tanashi,Tokyo 188,Japan } \\
\end{small}

\vspace{25pt}
\underline{ABSTRACT}
\end{center}

\vspace{10pt}
\begin{small}
The analogue of the string equation which specifies
the partition function of $c=1$ string with a
compactification radius
$\beta \in \mbox{$\bf{Z}$}_{\geq 1} $ is described in the
framework of the Toda lattice hierarchy.

\end{small}

\newpage

       Recently much attention has been paid to the understanding of
$c=1$ string theory from the view of integrable hierarchy.
In paticular the tachyon dynamics of the theory has been described
in the framework of the  (dispersionless) Toda lattice hierarchy
\cite{DMP},\cite{DTL},\cite{TEK}.
In spite of these developments our understanding of
the nonperturbative aspects for $c=1$ string still
has a gap from that of non-critical string theory.
In the non-critical string theory the full partition function is the
$\tau$ function of Kadomtsev-Petriashivil (KP) hierarchy specified
by the solution of string equation \cite{D} :
$\mbox{[}~P,~Q~\mbox{]}=1$ where $P,Q$ are differential operators.
These pairs can be given in terms of the
Lax and Orlov operators of KP hierarchy \cite{AM}.
On the other hand, even in the framework of the Toda lattice
(TL) hierarchy,
the analogue of the string equation at
$c=1$ has not been clarified yet.
In this letter, by utilizing the concepts of these integrable
hierarchies, we try to obtain this nonperturbative counterpart
which characterizes
the generating function for the tachyon correlation functions of
$c=1$ string with a compactification radius
$\beta  \in \mbox{$\bf{Z}$}_{\geq 1}$.

\section{$c=1$ string}

               The generating function $\cal F$ for
the tachyon correlation functions of
$c=1$ string with a compactification radius
$\beta \in \mbox{$\bf{Z}$}_{\geq 1}$ is described
in terms of free fermion system:
$\psi(z)=$  $\mbox{$\sum_{l}$} \psi_{l} z^{-l}$,
$\psi^{*}(z)=$ $ \mbox{$\sum_{l}$} \psi_{l}^{*} z^{-l-1}$.
The partition function is given  by
\begin{eqnarray}
&&
\exp \{  \mbox{$\cal{F}$} (t,\bar{t},\hbar) \}
      = ~~<0|e^{H(\frac{t}{\hbar})}g_0
               e^{-\overline{H}(\frac{\bar{t}}{\hbar})} |0>,
\label{Z}
\end{eqnarray}
where
$t=(t_1,t_2,\cdots)$ and $\bar{t}=(\bar{t}_1,\bar{t}_2,\cdots)$.
$t_k$ and $ \bar{t}_k~~ (k\in \mbox{$\bf{Z}$}_{\geq 1}) $
are parameters coupled to the
tachyons with momentum $\frac{k}{\beta}$ and $ \frac{-k}{\beta}$ respectively.
$"1/\hbar"$ will play the role of cosmological constant of this string
theory, that is,
$\mbox{$\cal{F}$}(t,\bar{t},\hbar)=
\sum_{g \geq 0}\hbar^{2g-2}\mbox{$\cal{F}$}_g(t,\bar{t})$,
where
$\mbox{$\cal{F}$}_g$
is the free energy on the 2 surface of genus $g$.
$g_{0}$ is an element of $GL(\infty)$ \cite{DMP},\cite{MPR}:
\begin{eqnarray}
&&
  g_0=~\exp \{ \sum_l
         \alpha_l :\psi_l\psi_{-l}^*: \} \nonumber\\
&&
   e^{\alpha_{l}}=\hbar^{-\frac{l-\frac{1}{2}}{\beta}}
         \frac{\Gamma(
                 \frac{1}{2}+\frac{1}{\hbar}
                       -\frac{l-\frac{1}{2}}{\beta})}
               {\Gamma(
                     \frac{1}{2}+\frac{1}{\hbar})}.
\label{data 1}
\end{eqnarray}
$H(t)$ and $\overline{H}(\bar{t})$ are respectively
the positive and negative modes of a
$U(1)$ current $J(z)=:\psi\psi^*:(z)$, that is,
\begin{eqnarray*}
H(t)= \mbox{$\sum_{k \geq 1}$}t_kJ_k~~,~~\overline{H}(\bar{t})
= \mbox{$\sum_{k \geq 1}$} \bar{t}_kJ_{-k}.
\end{eqnarray*}
The fermion vacuum $|n>$ is introduced  by the conditions :
\begin{eqnarray*}
\psi_k^*|n>&=&0 ~~~\mbox{for}~~k\geq n,\nonumber \\
\psi_k|n>&=&0~~~\mbox{for}~~k\geq 1-n.
\end{eqnarray*}

           In order to study the partition function
$e^{\cal{F}}$ (\ref{Z})
in the framework of the TL hierarchy we shall
introduce the quantities :
\begin{equation}
\tau(n:t,\bar{t})
=~<n|e^{H(\frac{t}{\hbar})}g_0
e^{-\overline{H}(\frac{\bar{t}}{\hbar})}
|n>.\label{tau}
\end{equation}
These guantities are precisely $\tau$-function of
the TL hierarchy \cite{UT,Take1}.
Since, by fixing $t=(t_1,t_2,\cdots)$
($\bar{t}=(\bar{t}_1,\bar{t}_2,\cdots)$),
$\tau(n;t,\bar{t})$ (\ref{tau})
can be considered as a special solution of the KP hierarchy which time
variables are $\bar{t}$ ($t$),
we shall begin by studying
$\tau(n;t,\bar{t})$ (\ref{tau})
from the view of the KP hierarchy  \cite{SS,DKJM}.

\section{$t$ as KP times}

  By introducing the Miwa variables
$\lambda=(\lambda_1,\cdots , \lambda_N)$
through the relation :
\begin{eqnarray*}
\frac{t_k}{\hbar}
=\frac{1}{k} \mbox{$\sum$}_{i=1}^N \lambda_i ^{-k}
{}~~~\mbox{for}~^\forall k
\in \mbox{$\bf{Z}$}_{\geq 1},
\end{eqnarray*}
and then utilizing the bozonization rule of free fermion
$\psi(z)$ and $\psi^*(z)$, one can rewrite $\tau(n)(\ref{tau})$ into
\begin{eqnarray*}
\tau(n;t,\bar{t}) &=&
\frac{\mbox{const.}}{\Delta(\lambda)}
\det \left|
       ~~\lambda_j^n
       <n|
             \psi_{i-n}\psi^*(\lambda_j)
                 g_0e^{-\overline{H}(\frac{\bar{t}}{\hbar})}
                          |n>~~
                         \right|_{1\leq i,j \leq N,}\nonumber\\
&=&
\mbox{const.}
       \frac{ (\prod_1^N \lambda_i)^n}{\Delta(\lambda)}
                    \det \left|
               ~~\widehat{\phi}_{i-n}(\lambda_j;\frac{\bar{t}}{\hbar})~~
                          \right|_{1\leq i,j \leq N} .
\end{eqnarray*}
$\Delta(\lambda)$ is the Vandermonde's determinant:
$\Delta(\lambda)= \prod_{i>j}^N(\lambda_i-\lambda_j)$.
$\widehat{\phi}_i(\lambda)$, which will play
an important role in the subsequent
discussion, has the form :
\begin{eqnarray*}
\widehat{\phi}_i(\lambda;\frac{\bar{t}}{\hbar})
=\sum_{l \geq 0}
\lambda^{i-l-1}
\frac{e^{\alpha_{i-l}}}{e^{\alpha_i}}
P_l(-\frac{\bar{t}}{\hbar}),
\end{eqnarray*}
where $P_l(\bar{t})$ is a Schur polynomial which
generating function is
\begin{eqnarray*}
\mbox{$\sum_{l\geq 0}$} P_l(\bar{t}) \lambda^l
=e^{\eta (\bar{t},\lambda)}~~,~~ \eta(\bar{t},
\lambda)=\mbox{$\sum_{l\geq 1}$} \bar{t}_l\lambda^l.
\end{eqnarray*}

                       One can consider $\tau(n;t,\bar{t})$ (\ref{tau})
as a $\tau$-function of the  KP hierarchy by
fixing $\bar{t}=(\bar{t}_1,\bar{t}_2,\cdots )$.
The corresponding point
in the Universal Grassmann manifold (UGM)
\cite{SS,SN} will be described in terms of
these $\phi_i(x;\frac{\bar{t}}{\hbar})$ or
the Laplace transtorms of
$\widehat{\phi}_i(\lambda;\frac{\bar{t}}{\hbar})$ :
\begin{equation}
\phi_i(x;\frac{\bar{t}}{\hbar})=
\int d \lambda
e^\frac{x \lambda}{\hbar}
\widehat{\phi}_i(\lambda;\frac{\bar{t}}{\hbar}).\label{base2}
\end{equation}

Let us define a vector space $V(n;\bar{t})$ (a point of the UGM) as
\begin{equation}
V(n;\bar{t})
=\bigoplus_{i\geq 1}
\mbox{$\bf{C}$}
\phi_{i-n}(x;\frac{\bar{t}}{\hbar}).
\label{Vn}
\end{equation}
Notice that the set
$\{ V(n;\bar{t}) \}_{n\in \bf{Z}}$ defines a flag of the UGM.
Namely it satisfies the conditions :
\begin{eqnarray*}
\bullet && \cdots \subset V(n;\bar{t})
      \subset V(n+1;\bar{t}) \subset  \cdots.  \nonumber\\
\bullet && \dim_{\bf{C}} \frac{V(n+1;\bar{t})}{V(n;\bar{t})} =1.
\end{eqnarray*}

            We can also introduce
$W(n;\hbar\partial_x)$, the wave operator of
KP hierarchy which corresponds to $\tau(n)$ (\ref{tau}).
This wave operator has the form :
\begin{eqnarray}
W(n;\hbar\partial_x)
&=&\sum_{j\geq 0}
u^{(\infty)}_{j,\alpha=\frac{1}{2}}(n)
(\hbar \partial_x)^{-j+n},
\label{wave1}
\end{eqnarray}
where $u^{(\infty)}_{j,\alpha=\frac{1}{2}}(n)
=u^{(\infty)}_{j,\alpha=\frac{1}{2}}(n;t,\bar{t})$.
$t=(t_1,t_2,\cdots)$ is  playing  the role of KP times
and especially we now identify
$t_1=x$.
The parameter $\alpha$ will play a gauge parameter in our
description of the $TL$ hierarchy.
The relation between the vector space $V(n;\bar{t})$ (\ref{Vn})
and the wave operator $W(n;\hbar\partial_x)$ (\ref{wave1})
is given by the theorem :
\newtheorem{th}{Theorem}
\begin{th}{\cite{SN}}
\begin{equation}
V(n;\bar{t})=W_0(n;\hbar\partial_x)^{-1} V_{\emptyset}, \label{TH1}
\end{equation}
where
\begin{eqnarray*}
W_0(n;\hbar\partial_x)=W(n;\hbar\partial_x)|_{t=(x,0,0....)}.
\end{eqnarray*}
\end{th}
$V_{\emptyset}$ is the point of the UGM which corresponds to the fermionic
vacuum $|0>$.
\footnote{By introducing the bases $\{e_i(x)\}$,
$V_{\emptyset}=\bigoplus_{i \geq 1}\mbox{$\bf{C}$}e_{-i}(x)$ holds.
Notice that, in our notation,
$\hbar \partial_x$ and $x$ act on these bases as
$\hbar \partial_x e_i(x)=e_{i-1}(x)$,
$xe_i(x)=\hbar(i+1)e_{i+1}(x)$.}

                            Owing to the above theorem we can characterize
the wave operator $W(n)\equiv W(n;\hbar\partial_x)$
(\ref{wave1}) from the study of the point $V(n,\bar{t})$ (\ref{Vn}).
In paticular one can see
$\phi_{i} (x;\frac{\bar{t}}{\hbar})$ (\ref{base2})
satisfies the following equations:
For $\mbox{}^{\forall} k \in \mbox{$\bf{Z}$}_{\geq 1}$,
\begin{eqnarray}
\mbox{(i)}&&~~
D_{\beta}(x;\hbar\partial_x)^k \phi_i (x; \frac{\bar{t}}{\hbar})
     ~=~ \hbar^k
\frac{\Gamma(k+\frac{1}{2}+\frac{1}{\hbar}-\frac{i-\frac{1}{2}}{\beta})}
           {\Gamma(\frac{1}{2}+\frac{1}{\hbar}-\frac{i-\frac{1}{2}}{\beta})}
   \phi_{i-k\beta}(x;\frac{\bar{t}}{\hbar}).
\nonumber \\
\mbox{(ii)}&&~~
-\hbar \frac{\partial}{\partial\bar{t}_{k\beta}}
               \phi_{i}(x;\frac{\bar{t}}{\hbar})
{}~=~(-\hbar
      \frac{\partial}{\partial\bar{t}_{\beta}})^k
            \phi_{i}(x;\frac{\bar{t}}{\hbar})
           \nonumber \\
&& ~~~~~~~~~~~ ~~~~~~~~~~~~~~=~\hbar^{k}
     \frac{ \Gamma(k+\frac{1}{2}+\frac{1}{\hbar}
                      - \frac{i-\frac{1}{2}} {\beta})}
          {\Gamma(\frac{1}{2}+\frac{1}{\hbar}-
                     \frac{i-\frac{1}{2}}{\beta})}
     \phi_{i-k\beta}(x;\frac{\bar{t}}{\hbar}).
\label{eq1}
\end{eqnarray}
The pseudo-differential operator
$D_{\beta}\equiv D_{\beta}(x;\hbar\partial_x) $
in (i) of (\ref{eq1}) is :
\begin{equation}
D_{\beta}(x;\hbar\partial_x)
{}~=~
\frac{1}{\beta}(\hbar\partial_x)^{1-\beta}x
    +\{1+ \frac{\hbar}{2}(1-\frac{1}{\beta})\}
           (\hbar\partial_x)^{-\beta}.
\label{D}
\end{equation}
Let us look into the content of the equation (ii) in (\ref{eq1}).
Introducing pseudo-differential operators
$\overline{P}_{k\beta}\equiv \overline{P}_{k\beta}
(x;\hbar\partial_x)$ ($k\in \mbox{$\bf{Z}$}_{\geq1}$) which satisfy
\begin{eqnarray*}
-\hbar \partial_{\bar{t}_{k\beta}}\xi ~=~ \overline{P}_{k\beta}
(x;\hbar\partial_x)\xi,
\end{eqnarray*}
for $\mbox{}^\forall\xi \in V =\bigcup_n V(n;\bar{t})$
($\equiv\bigoplus_i
\mbox{\bf{C}} \phi_i(x;\frac{\bar{t}}{\hbar})$),
the first line of (ii) in (\ref{eq1}) tells us the relation :
\begin{eqnarray*}
\overline{P}_{k\beta}
{}~=~
\underbrace{\overline{P}_\beta \cdot ~ \cdots ~ \cdot \overline{P}_\beta}_{k}.
\end{eqnarray*}
Therefore, from theorem 1, the evolution of
$W_0(n)\equiv W_0(x;\hbar \partial_x)$
with respect to $\bar{t}_{\beta k}$ has the form
\footnote{
$(~~)_{\pm}$ are introduced as projection operators
with respect to $\partial_x$;
\begin{eqnarray*}
(\partial_x^m)_{-}= \partial_{x}^m-(\partial_x^m)_+,~~~~ (\partial_x^m)_+
=\left\{\begin{array}{ll}
\partial_x^m & m\geq 0\\
0            & m<-1
\end{array}\right.
\end{eqnarray*}} :
\begin{equation}
\hbar\partial_{\bar{t}_{k\beta}}W_0(n)~=~
(W_0(n)\overline{P}_{\beta}^kW_0(n)^{-1})_-
W_0(n). \label{eq3}
\end{equation}

         At this stage we shall turn to (i) in (\ref{eq1}).
By combining
(i) with (ii), we also obtain the relation :
\begin{equation}
\overline{P}_{\beta} ~=~ D_{\beta}.
\label{eq5}
\end{equation}
This can be rephrased by using the Orlov formulation of KP
hierarchy \cite{Or}.
Let us introduce  the Lax operator
$L^{KP}(n)$ $\left(\equiv L^{KP}(n;\hbar\partial_x) \right)$
and the Orlov operator
$M^{KP}(n)$ $\left(\equiv M^{KP}(n;\hbar\partial_x) \right)$
as
\begin{eqnarray*}
L^{KP}(n;\hbar\partial_x) &=& W(n) \hbar\partial_x W(n)^{-1},
\nonumber\\
M^{KP}(n;\hbar\partial_x) &=& W(n)
       \left( \mbox{$\sum_{l\geq 1}$}
            lt_l(\hbar\partial_x)^{l-1} \right)
                        W(n)^{-1}.
\end{eqnarray*}
In terms of these operators  the relation
(\ref{eq5}) becomes :
\begin{equation}
W_0(n)\overline{P}_{\beta}W(n)_0^{-1}~=~
F(L_0^{KP}(n), M_0^{KP}(n)) ,
\label{eq7}
\end{equation}
where
\begin{equation}
F(x,y)= \frac{1}{\beta} x^{1-\beta}y +\{1+\frac{\hbar}{2}
(1-\frac{1}{\beta})\}x^{-\beta} .
\end{equation}
$L_0^{KP}(n)$ and $M_0^{KP}(n)$ are those  evaluated
at the initial time $t=(x,0,0,\cdots)$, that is,
\begin{eqnarray*}
L_0^{KP}(n) = L^{KP}(n)|_{t=(x,0,0,\cdots)} ,~~
M_0^{KP}(n) = M^{KP}(n) |_{t=(x,0,0,\cdots)}.
\end{eqnarray*}

        Finally we will  return to the equation (\ref{eq3}).
Though $\bar{t}=(\bar{t}_1,\bar{t}_2,\cdots)$ can not be considered
as KP times in our present approach, it is possible to define their flows as
\begin{eqnarray}
\hbar \partial_{\bar{t}_m} W(n) &=&
\overline{B}_m^{KP}(n)W(n),
\nonumber\\
\overline{B}_m^{KP}(n) &=& \sum_{l\geq1} P_l(-\hbar\tilde{\partial}_t)
(\hbar\partial_{\bar{t}_m})\ln \tau(n)L^{KP}(n)^{-l},
\label{equation10}
\end{eqnarray}
where $\tilde{\partial}_t =(\partial_1,\frac{1}{2}\partial_{t_2},\cdots)$.
Hence the equation (\ref{eq3}) means
\begin{equation}
(W_0(n)\overline{P}^k_\beta W_0(n)^{-1})_-
{}~=~ \overline{B}^{KP}_{k\beta}(n)|_{t=
(x,0,\cdots)}~~~~\mbox{for}~^\forall k \in \mbox{$\bf{Z}$}_{\geq 1}.
\label{eq11}
\end{equation}

\section{$\bar{t}$ as KP times}

       Nextly we will  consider $\tau(n;t,\bar{t})$ (\ref{tau})
as a $\tau$ function of
the KP hierarchy with respect to time variables
$\bar{t}=(\bar{t}_1,\bar{t}_2,\cdots)$.

     Introducting the Miwa variables
$\lambda=(\lambda_1,\cdots,\lambda_N) $ though the
relation :
\begin{eqnarray*}
\frac{\bar{t}_k}{\hbar}~=~
\frac{1}{k}\sum^N_{i=1} \lambda_i^{-k}~~~\mbox{for}~~^\forall
k \in \mbox{$\bf{Z}$}_{\geq 1},
\end{eqnarray*}
the $\tau$ function $\tau(n+1)$ (\ref{tau}) can be written into
\begin{eqnarray*}
\tau(n+1;t,\bar{t})
{}~=~ \frac{\mbox{const.}}
         {( \mbox{$\prod^N_{i=1}$}\lambda_i)^{n+1}\Delta(\lambda)}
    \det \left|~~
       \widehat{\overline{\phi}}_{i+n+1}
       (\lambda_j; \frac{t}{\hbar})~~\right|_{1\leq i,j\leq N},
\end{eqnarray*}
where
\begin{eqnarray*}
\widehat{\overline{\phi}}_{i}(\lambda; \frac{t}{\hbar})
{}~=~\sum_{l\geq 0} \lambda^{i-l-1}
    \frac{e^{\alpha_{1-i}}}{e^{\alpha_{1+l-i}}} P_l(-\frac{t}{\hbar}).
\end{eqnarray*}
One can consider $\tau(n+1)$ (\ref{tau}) as a $\tau$ function of
the KP hierarchy with KP times
$\bar{t}$.
The corresponding point in the UGM which we will
denote as $\overline{V}(n;t)$
can be described in terms of
$\overline{\phi}_{i}(x;\frac{t}{\hbar})
  ~=~ \int d\lambda e^{\frac{x \lambda}{\hbar}}
\widehat{\overline{\phi}}_i
(\lambda;\frac{t}{\hbar})~$ :
\begin{equation}
\overline{V}(n;t)
= \bigoplus_{i\geq 1} \mbox{$\bf{C}$}
\overline{\phi}_{i+n+1}(x;\frac{t}{\hbar}).
\label{barVn}
\end{equation}
Note that, as we see in the previous section, the set
$\{\bar{V}(n;t)\}_{n \in \mbox{\bf{Z}}}$ defines a flag in the
UGM.

               $\overline{W}(n;\hbar\partial_x)$,
the wave operator of KP hierarechy which corresponds to
$\overline{V}(n;t)$ (\ref{barVn}), will be realized as
\begin{eqnarray}
\overline{W}(n;\hbar\partial_x)
{}~=~ \mbox{$\sum$}_{j\geq 0}u^{(0)}_{j,\alpha=-\frac{1}{2}}(n)
(\hbar\partial_x)^{-j-n-1},
\label{wave2}
\end{eqnarray}
where $u^{(0)}_{j,\alpha=-\frac{1}{2}}(n)
=u^{(0)}_{j,\alpha=-\frac{1}{2}}(n;t,\bar{t})$ and
we specify $x=\bar{t}_1$.
The relation between $\overline{V}(n;t)$ and
$\overline{W}(n;\hbar\partial_x)$ is
\begin{eqnarray*}
\overline{V}(n;t) ~=~ \overline{W}_0(n;\partial_x)^{-1}V_{\emptyset},
\end{eqnarray*}
where
\begin{eqnarray*}
\overline{W}_0(n;\hbar\partial_x)
{}~=~\overline{W}(n;\hbar\partial_x)|_{\bar{t}=(x,0,0, \cdots )}.
\end{eqnarray*}

            Since $\overline{\phi}_i(x;\frac{t}{\hbar})$ satisfies
the similar equations as those
presented in (\ref{eq1})  we can follow the same steps as in the
previous section.
Especially the analogue of the relation (\ref{eq5}) is
\begin{equation}
P_\beta ~=~ - \overline{D}_\beta .
\label{eq33}
\end{equation}
$\overline{D}_\beta=\overline{D}_\beta (x;\hbar\partial_x)$ is
the pseudo-differential operator with the following form
\begin{eqnarray*}
\overline{D}_\beta (x;\hbar\partial_x)
=\frac{1}{\beta}(\hbar\partial_x)^{1-\beta}x
  -\{1-\frac{\hbar}{2}(1-\frac{1}{\beta})\}
       (\hbar\partial_x)^{-\beta}.
\end{eqnarray*}
On the other hand the pseudo-differential operator
$P_\beta=P_\beta(x;\hbar\partial_x)$
is given by the conditions :
\begin{eqnarray*}
-\hbar \partial_{ t_\beta} \bar{\xi} ~=~
 P_\beta(x,\hbar\partial_x)\bar{\xi},
\end{eqnarray*}
for $\mbox{}^\forall \bar{\xi} \in \overline{V}
= \bigcup_{n} \overline{V}(n;t)$.

                        Introducing the Lax operator
$\overline{L}^{KP}(n)
\left( \equiv \overline{L}^{KP}(n;\hbar\partial_x) \right)~$
and the Orlov operator
$\overline{M}^{KP}(n)
\left( \equiv \overline{M}^{KP}(n;\hbar\partial_x) \right)~$ as
\begin{eqnarray*}
&&\overline{L}^{KP}(n;\hbar\partial_x)
 ~=~\overline{W}(n)(\hbar\partial_x)\overline{W}(n)^{-1},
\nonumber\\
&&\overline{M}^{KP}(n;\hbar\partial_x)
{}~=~\overline{W}(n)
   \left( \sum_{l\geq1}l\bar{t}_l(\hbar\partial_x)^{l-1} \right)
 \overline{W}(n)^{-1},
\end{eqnarray*}
then we can rewrite the relation (\ref{eq33}) into :
\begin{equation}
\overline{W}_0(n) P_\beta^k\overline{W}_0(n)^{-1}
{}~=~
- G(\overline{L}_0^{KP}(n),\overline{M}_0^{KP}(n)) ,
\label{eq77}
\end{equation}
where
\begin{equation}
G(x,y) ~=~
\frac{1}{\beta}x^{1-\beta}x +\{-1+\frac{\hbar}{2}
(1-\frac{1}{\beta}) \} x^{-\beta}.
\label{eq88}
\end{equation}
This is the analogue of the relation
(\ref{eq7}).

                      Because the analogue of the equation
(\ref{eq3}) is
\begin{eqnarray*}
\hbar \partial_{ t_{k\beta}} \overline{W}_0 (n)
{}~=~ (\overline{W}_0(n) P_{\beta}^k
\overline{W}_0(n)^{-1})_{-}\overline{W}_0(n) ,
\end{eqnarray*}
we can conclude that
$\left( \overline{W}_0(n) P_{\beta}^k \overline{W}_0(n)^{-1} \right)_-$
($k \in \mbox{$\bf{Z}$}_{\geq 1}$) has the  expression :
\begin{eqnarray}
 \left(
\overline{W}_0(n) P_{\beta}^k \overline{W}_0(n)^{-1}
\right)_-
{}~= ~
B_{k \beta}^{KP}(n) |_{\bar{t}=(x,0, \cdots)},
{}~~~~\mbox{for}~^{\forall}k \in
\mbox{$\bf{Z}$}_{\geq 1},
\label{eq555}
\end{eqnarray}
where
\begin{eqnarray}
B_{k \beta}^{KP}(n)~=~
\mbox{$\sum$}_{j \geq 1}P_j(-\hbar \tilde{\partial}_{\bar{t}})
           (\hbar \partial_{\bar{t}_{k \beta}}) \ln \tau (n+1)
                   \overline{L}^{KP}(n)^{-j}.
\label{eq55}
\end{eqnarray}
This corresponds to the equation (\ref{eq11}).

\section{TL hierarchy}

    In this section we will give equivalent expressions
for the relations (\ref{eq7})
and (\ref{eq77}) in the terminology of the TL hierarchy.
These conditions specify
$\tau(n)$ (\ref{tau}).
For this purpose
let us begin by reviewing the TL hierarchy along the line of those in
\cite{UT,Take2}.

                The wave operators of TL hierarchy in $\alpha$-
gauge which we will denote as
$W_\alpha^{(^{\infty}_{0})}(n;e^{\partial_n})$
 are given by
\begin{eqnarray}
W_\alpha^{(^{\infty}_{0})}(n;e^{\partial_n})
= \sum_{j\geq 0}
u_{j,\alpha}^{(^{\infty}_{0})}(n)
e^{\mp j \partial_n},
\label{wave3}
\end{eqnarray}
where
$u_{j,\alpha}^{(^{\infty}_{0})}(n)=
u_{j,\alpha}^{(^{\infty}_{0})}(n;t,\bar{t})$,
and $"e^{\partial_n}"$ is a difference operator with respect to $n$.
The coefficents
$u_{j,\alpha}^{(^{\infty}_{0})}(n)$ can be described in terms of
a $\tau$ function of the TL hierarchy :
\begin{equation}
u_{j,\alpha}^{(\infty)}(n)=
\frac{P_j(-\hbar \tilde{\partial}_t)\tau(n)}
       {\tau(n+1)^{\frac{1}{2}-\alpha}
                 \tau(n)^{\frac{1}{2}+\alpha}}~,~~
u_{j,\alpha}^{(0)}(n)=
\frac{P_j(-\hbar \tilde{\partial}_{\bar{t}})\tau(n+1)}
       {\tau(n+1)^{\frac{1}{2}-\alpha}
                 \tau(n)^{\frac{1}{2}+\alpha}}~.
\label{upot}
\end{equation}

                 The time evolutions of these wave operators
are
\begin{eqnarray*}
\hbar\partial_{t_m}W_\alpha^{(\infty)}(n)&=&
   B^{TL}_{m,\alpha}(n)W_\alpha^{(\infty)}(n)
      -W_\alpha^{(\infty)}(n)e^{m\partial_n},
\\
\hbar\partial_{\bar{t}_m}W_\alpha^{(\infty)}(n)&=&
    \overline{B}^{TL}_{m,\alpha}(n)
           W_\alpha^{(\infty)}(n),
\\
\hbar\partial_{t_m}W_\alpha^{(0)}(n)&=&
      B^{TL}_{m,\alpha}(n)W_\alpha^{(0)}(n),
\\
\hbar\partial_{\bar{t}_m}W_\alpha^{(0)}(n)&=&
     \overline{B}^{TL}_{m,\alpha}(n)W_\alpha^{(0)}(n)
         -W_\alpha^{(0)}(n)e^{-m\partial_n},
\end{eqnarray*}
where the difference operators
$B^{TL}_{m,\alpha}(n)$ and
$\overline{B}^{TL}_{m,\alpha}(n)$
are
\footnote{
Notice that $(~~)_{\pm,0}$ are projection operators with respect to
$e^{\partial_n}$;
\begin{eqnarray*}
   (e^{m\partial_n})_+
     &=&
       \left\{\begin{array}{ll}
                 e^{m\partial_n} & m\geq 1\\
                               0 & m\leq 0\\
                                  \end{array}\right.,
\\
    (e^{m\partial_n})_0
       &=& \delta_{m,~0},~~
    (e^{m\partial_n})_-
             =e^{m\partial_n}
                 - (e^{m\partial n})_+
                      -(e^{m\partial_n})_0.
\end{eqnarray*}}
\begin{eqnarray*}
B^{TL}_{m,\alpha}(n)&=&
    (L^{TL}_{\alpha}(n)^m)_+
        +(\frac{1}{2}+\alpha)
           (L^{TL}_{\alpha}(n)^m)_0,
\\
\overline{B}^{TL}_{m,\alpha}(n)&=&
    (\overline{L}^{TL}_{\alpha}(n)^m)_-
        +(\frac{1}{2}-\alpha)
           (\overline{L}^{TL}_{\alpha}(n)^m)_0,
\end{eqnarray*}
with the following Lax operators :
\begin{eqnarray*}
L^{TL}_{\alpha}(n)&=&
   W_\alpha^{(\infty)}(n)e^{\partial_n}W_\alpha^{(\infty)}(n)^{-1},
 \\
\overline{L}^{TL}_{\alpha}(n)&=&
   W_\alpha^{(0)}(n)e^{-\partial_n}W_\alpha^{(0)}(n)^{-1}.
\end{eqnarray*}


       The wave functions of TL hierarchy are defined by
\begin{eqnarray*}
\psi^{(\infty)}_{\alpha}(n;\lambda)
    &=& W^{(\infty)}_{\alpha}(n)
            e^{\frac{1}{\hbar}\eta(t;\lambda)}\lambda^n~,
\\
\psi^{(0)}_{\alpha}(n;\lambda)
    &=& W^{(0)}_{\alpha}(n)
            e^{\frac{1}{\hbar}\eta(\bar{t};\frac{1}{\lambda})}\lambda^n~,
\end{eqnarray*}
on which the Lax operators act as
\begin{eqnarray*}
L^{TL}_{\alpha}(n)\psi^{(\infty)}_{\alpha}(n;\lambda)
 &=& \lambda \psi^{(\infty)}_{\alpha}(n;\lambda),~~
M^{TL}_{\alpha}(n)\psi^{(\infty)}_{\alpha}(n;\lambda)
 =\hbar \partial_{\lambda} \psi^{(\infty)}_{\alpha}(n;\lambda), \nonumber \\
\overline{L}^{TL}_{\alpha}(n)\psi^{(0)}_{\alpha}(n;\lambda)
 &=& \frac{1}{\lambda} \psi^{(0)}_{\alpha}(n;\lambda),~~
\overline{M}^{TL}_{\alpha}(n)\psi^{(0)}_{\alpha}(n;\lambda)
 =-\hbar \lambda^2 \partial_{\lambda} \psi^{(0)}_{\alpha}(n;\lambda),
\end{eqnarray*}
where we also introduce the  Orlov operators
$M^{TL}_{\alpha}(n)$
$\left(\equiv
M^{TL}_{\alpha}(n;e^{\partial_n}) \right)$,
$\overline{M}^{TL}_{\alpha}(n)$  $\left(\equiv
\overline{M}^{TL}_{\alpha}(n; e^{\partial_n}) \right)$ by
\begin{eqnarray*}
M^{TL}_{\alpha}(n)&=&
 W_{\alpha}^{(\infty)}(n)
   \left( \sum_{l \geq 1}
           lt_le^{(l-1)\partial_n}+\hbar n e^{-\partial_n} \right)
                    W_{\alpha}^{(\infty)}(n)^{-1}, \nonumber \\
\overline{M}^{TL}_{\alpha}(n)&=&
 W_{\alpha}^{(0)}(n)
   \left( \sum_{l \geq 1}
           l\bar{t}_le^{-(l-1)\partial_n}-\hbar n e^{\partial_n} \right)
                    W_{\alpha}^{(0)}(n)^{-1}.
\end{eqnarray*}

{}~

                 With the above brief review of the
TL hierarchy we shall return
to our specific solution of the TL hierarchy (\ref{tau}).
We first notice that, in $\alpha=\frac{1}{2}$ gauge,one can expand
$\overline{B}^{TL}_{k \beta,~\alpha=\frac{1}{2}}(n)$
by negative powers of
$L^{TL}_{\alpha=\frac{1}{2}}$ :
\begin{eqnarray}
\overline{B}_{k \beta, \alpha=\frac{1}{2}}^{TL}(n)
&\equiv &
\left( \overline{L}^{TL}_{\alpha=\frac{1}{2}}(n)^{k \beta} \right)_-
\nonumber \\
&=&
\sum_{l\geq1} P_l(-\hbar\tilde{\partial}_t)
(\hbar\partial_{\bar{t}_{k \beta}})\ln \tau(n)
L^{TL}_{\alpha=\frac{1}{2}}(n)^{-l}.
\label{BTLEXP1}
\end{eqnarray}
By comparing the equation (\ref{BTLEXP1}) with
the equation (\ref{equation10})
one can see
$W(n)\overline{P}_{\beta}W(n)^{-1}$
plays the same role as
$\overline{L}^{TL}_{\alpha=\frac{1}{2}}(n)^{\beta}$.
In paticular the actions of these two operators on the wave function
$\psi^{(\infty)}_{\alpha=\frac{1}{2}}(n;\lambda)$ should be same :
\begin{eqnarray}
W(n)\overline{P}_{\beta}W(n)^{-1}
\psi^{(\infty)}_{\alpha=\frac{1}{2}}(n;\lambda)
=\overline{L}^{TL}_{\alpha=\frac{1}{2}}(n)^{\beta}
\psi^{(\infty)}_{\alpha=\frac{1}{2}}(n;\lambda).
\label{KPTL1}
\end{eqnarray}
Note that  the Lax operators
$L^{KP}(n)$,
   $L^{TL}_{\alpha=\frac{1}{2}}(n)$
and the Orlov operators
$M^{KP}(n)$,
$M^{TL}_{\alpha=\frac{1}{2}}(n)$ act on the wave function
$\psi^{(\infty)}_{\alpha=\frac{1}{2}}(n;\lambda)$ as
\begin{eqnarray}
L^{KP}(n)
      \psi^{(\infty)}_{\alpha=\frac{1}{2}}(n;\lambda)
  &=&
     L^{TL}_{\alpha=\frac{1}{2}}(n)
        \psi^{(\infty)}_{\alpha=\frac{1}{2}}(n;\lambda)
\nonumber \\
  &=&
     \lambda \psi^{(\infty)}_{\alpha=\frac{1}{2}}(n;\lambda)
\nonumber \\
M^{KP}(n)\psi^{(\infty)}_{\alpha=\frac{1}{2}}(n;\lambda)
 &=&
     M^{TL}_{\alpha=\frac{1}{2}}(n)
\psi^{(\infty)}_{\alpha=\frac{1}{2}}(n;\lambda)
\nonumber \\
 &=&
      \hbar \partial_{\lambda}
        \psi^{(\infty)}_{\alpha=\frac{1}{2}}(n;\lambda).
\label{KPTL2}
\end{eqnarray}
Hence from the above equalities (\ref{KPTL1}) and (\ref{KPTL2})
we can express the relation (\ref{eq7}) in the terminology of the
TL hierarchy :
\begin{eqnarray}
\overline{L}^{TL}_{\alpha=\frac{1}{2}}(n)^{\beta}
{}~=~
F(L^{TL}_{\alpha=\frac{1}{2}}(n),
    M^{TL}_{\alpha=\frac{1}{2}}(n)).
\label{TL1}
\end{eqnarray}
This condition should be independent of the gauge.
{}From the explicit forms
(\ref{upot}) of $W_{\alpha}^{(^{\infty}_0)}(n)$
their gauge transforms are simple :
$W_{\alpha+\gamma}^{(^{\infty}_{0})}(n)=
 f(n;\gamma)W_{\alpha}^{(^{\infty}_{0})}(n)$. The equation
(\ref{TL1}) is preserved under the gauge transformation.
Therefore the relation (\ref{eq7}) is eqivalent to
\footnote{Here we abbreviate the gauge parameter $\alpha $.}
\begin{eqnarray}
\overline{L}^{TL}(n)^{\beta}=
\frac{1}{\beta}
  \left(L^{TL}(n) \right)^{-\beta}
    \{ M^{TL}(n)+\left( \beta
           +\frac{\hbar}{2}(1+\beta)\right)L^{TL}(n)^{-1}\}
          L^{TL}(n).
\label{TL2}
\end{eqnarray}
This is one of two relations which will specify
$e^{\cal{F}}$ (\ref{Z}) in the framework of TL hierarchy.

          We shall try to obtain the another equation
which is characteristic of
$c=1$ string with a compactification radius
$\beta \in \mbox{\bf{Z}}_{\geq 1}$.
For this purpose we will examine $\alpha =-\frac{1}{2}$ gauge
of TL hierarchy.
In $\alpha=-\frac{1}{2}$ gauge,
$ B^{TL}_{k \beta,\alpha=-\frac{1}{2}}(n)$ can be expanded by the
negative powers of $\overline{L}^{TL}_{\alpha=-\frac{1}{2}}(n)$ :
\begin{eqnarray}
 B_{k \beta, \alpha=-\frac{1}{2}}^{TL}(n)
&\equiv &
\left( L^{TL}_{\alpha=-\frac{1}{2}}(n)^{k \beta} \right)_+
\nonumber \\
&=&
\sum_{l\geq1} P_l(-\hbar\tilde{\partial}_{\bar{t}})
(\hbar\partial_{t_{k \beta}})\ln \tau(n+1)
\overline{L}^{TL}_{\alpha=-\frac{1}{2}}(n)^{-l}.
\label{BTLEXP2}
\end{eqnarray}
By comparing the equation (\ref{BTLEXP2}) with the equation (\ref{eq55})
one can see that
$\overline{W}(n)P_{\beta}\overline{W}(n)^{-1}$ corresponds to
$L^{TL}_{\alpha=-\frac{1}{2}}(n)^{\beta}$ :
\begin{eqnarray*}
\overline{W}(n)P_{\beta}\overline{W}(n)^{-1}
\psi^{(0)}_{\alpha=-\frac{1}{2}}(n;\lambda)
=L^{TL}_{\alpha=-\frac{1}{2}}(n)^{\beta}
\psi^{(0)}_{\alpha=-\frac{1}{2}}(n;\lambda).
\end{eqnarray*}
Since the Lax operators
$\overline{L}^{KP}(n)$,
$\overline{L}^{TL}_{\alpha=-\frac{1}{2}}(n)$
and the Orlov operators
$\overline{M}^{KP}(n)$,
$\overline{M}^{TL}_{\alpha=-\frac{1}{2}}(n)$
act on the wave function
$\psi^{(0)}_{\alpha=-\frac{1}{2}}(n;\lambda)$ as
\begin{eqnarray*}
\overline{L}^{KP}(n)
      \psi^{(0)}_{\alpha=-\frac{1}{2}}(n;\lambda)
&=&
     \overline{L}^{TL}_{\alpha=-\frac{1}{2}}(n)
        \psi^{(0)}_{\alpha=-\frac{1}{2}}(n;\lambda)
\nonumber \\
&=&
     \frac{1}{\lambda} \psi^{(0)}_{\alpha=
                 -\frac{1}{2}}(n;\lambda),
\nonumber \\
\left(
\overline{M}^{KP}(n)
+\hbar \overline{L}^{KP}(n)^{-1} \right)
\psi^{(0)}_{\alpha=-\frac{1}{2}}(n;\lambda)
&=&
     \overline{M}^{TL}_{\alpha=-\frac{1}{2}}(n)
\psi^{(0)}_{\alpha=-\frac{1}{2}}(n;\lambda)
\nonumber \\
&=&
    - \hbar \lambda^2 \partial_{\lambda}
        \psi^{(0)}_{\alpha=-\frac{1}{2}}(n;\lambda),
\end{eqnarray*}
the relation (\ref{eq77}) can be rephrased into
\begin{eqnarray*}
L^{TL}_{\alpha=-\frac{1}{2}}(n)^{\beta}
{}~=~ -
G(\overline{L}^{TL}_{\alpha=-\frac{1}{2}}(n),
    \overline{M}^{TL}_{\alpha=-\frac{1}{2}}
           -\hbar \overline{L}^{TL}_{\alpha=
                            -\frac{1}{2}}(n)^{-1})  ,
\end{eqnarray*}
which turns out the following gauge invariant form :
\begin{eqnarray}
L^{TL}(n)^{\beta}=
-\frac{1}{\beta} \left( \overline{L}^{TL}(n)\right)^{-\beta}
   \{ \overline{M}^{TL}(n)-
             \left(
                   \beta+ \frac{\hbar}{2}(1-\beta) \right)
                 \overline{L}^{TL}(n)^{-1} \}
                        \overline{L}^{TL}(n).
\label{TL4}
\end{eqnarray}

{}~

              It is important to remark that the above obtained relations
(\ref{TL2}) and (\ref{TL4}) are consistent with the commutation
relations among the Lax and Orlov operators in the TL hierarchy :
\begin{eqnarray*}
\mbox{[}
L^{TL}(n)~,~M^{TL}(n) \mbox{]}~
  =~~\hbar,~~~~~\mbox{[}
\overline{L}^{TL}(n)~,~\overline{M}^{TL}(n) \mbox{]}~
  =~~\hbar.
\end{eqnarray*}
This consistency tells us
that one can regard the pair of these relations as
the  "twistor data" of TL hierarchy \cite{TT} which is associated
with the solution (\ref{tau}).
Thus the relations (\ref{TL2}) and (\ref{TL4}) characterize
the generating function for the tachyon amplitudes of
$c=1$ string with a compactification radius $\beta \in \mbox{\bf{Z}}_{\geq 1}$.

\section{Comment}

          Let us first comment on the nature of $\mbox{$\cal{F}$}_0$,
the genus $0$ contribution to $\mbox{$\cal{F}$}$.
This quantity may be described in terms of the $\tau$ function of the
dispersionless Toda hierarchy \cite{TT} which is the
$\hbar \rightarrow 0$ limit of the TL hierarchy with fixing
$s=\frac{n}{\hbar}$.
In this limit the Lax and Orlov operators of the TL hierarchy turn
to the counterparts of this dispersionless hierarchy :
\footnote{We follow the notation of the reference \cite{TT}.}
\begin{eqnarray*}
L(n) (\equiv L^{TL}(n)) \rightarrow \mbox{$\cal{L}$}(s),
&&
M(n) (\equiv M^{TL}(n)L^{TL}(n)) \rightarrow \mbox{$\cal{M}$}(s),
\nonumber \\
\widehat{L}(n) (\equiv \overline{L}^{TL}(n)^{-1}) \rightarrow
\widehat{\mbox{$\cal{L}$}}(s),
&&
\widehat{M}(n) (\equiv -\overline{M}^{TL}(n)\overline{L}^{TL}(n))
\rightarrow \widehat{\mbox{$\cal{M}$}}(s).
\end{eqnarray*}
In this limit the string equation (\ref{TL2}) and (\ref{TL4})
becomes as follows :
\begin{eqnarray*}
\widehat{\mbox{$\cal{L}$}}(s)^{-\beta}
=\frac{1}{\beta}\mbox{$\cal{L}$}(s)^{-\beta}(\mbox{$\cal{M}$}(s)+\beta),~~
\mbox{$\cal{L}$}(s)^{\beta}=
\frac{1}{\beta}\widehat{\mbox{$\cal{L}$}}(s)^{\beta}
(\widehat{\mbox{$\cal{M}$}}(s)+\beta).
\end{eqnarray*}
Notice that, without any effect on the physical quantities,
we can shift $s$, the zero-th time variable of this dispersionless
hierarchy, to $s-\beta$ in the definition of the Orlov operators
$\mbox{$\cal{M}$}(s)$ and $\widehat{\mbox{$\cal{M}$}}(s)$ \cite{TT}.
With this harmless shift we obtain
\begin{eqnarray}
\widehat{\mbox{$\cal{L}$}}(s)^{-\beta}=
\frac{1}{\beta}\mbox{$\cal{L}$}(s)^{-\beta}\mbox{$\cal{M}$}(s),~~
\mbox{$\cal{L}$}(s)^{\beta}=
\frac{1}{\beta}\widehat{\mbox{$\cal{L}$}}(s)^{\beta}
\widehat{\mbox{$\cal{M}$}}(s).
\label{0string}
\end{eqnarray}
In the case of $\beta=1$ this is precisely the proposed twistor data
appropriate to describe  the classical (genus 0) tachyon dynamics at
the self dual radius \cite{TEK}.

     In this classical limit the $\hbar-$dependent parts in (\ref{TL2})
and (\ref{TL4}) vanish. Neverthless this vanishing parts can be expected
to play an important role in the nonperturbative definition of the theory.
For an example,
in the case of $A_{N-1}$ topological string,
the partition function is the $\tau$ function of the
$N-$ reduced KP hierarchy
specified by the following form of the string equation
\cite{AM} :
\begin{eqnarray}
P = L^N,~~
Q = \frac{1}{N}\left(
               M-NL^N-\frac{N-1}{N}\hbar L^{-1}\right)
                        L^{1-N}.
\label{STRING}
\end{eqnarray}
The $\hbar-$dependent term in (\ref{STRING}) is crucial
for the matrix integral realization of this topological
string \cite{AM},\cite{NKNT}. And also,
from the analysis of the above equation (\ref{STRING})
some geometrical nature of the topological string
has been revealed through the "genus expansion" \cite{NKNT}.
Thus we can also expect that some geometry of $c=1$ string
theory appears from the study of the characteristic relations
(\ref{TL2}) and (\ref{TL4}).This study will be reported elsewhere.

{}~

{}~

      The author would like to thank Prof.T.Eguchi for several
discussions and  hospitality during his stay at Hongo,
Univ. of Tokyo.He would also like to thank Prof.M.Noumi
, Dr.T.Takebe for useful discussions and comments, and
Prof.K.Takasaki for the comments on the first version of this note.

\end{document}